\documentclass[aps,showpacs,superscriptaddress,preprint]{revtex4}%
\usepackage{amsfonts}
\usepackage{amsmath}
\usepackage{amssymb}
\usepackage{graphicx}%
\usepackage{supertabular}
\usepackage{longtable}
\providecommand{\U}[1]{\protect\rule{.1in}{.1in}}

\begin{document}
\title{Wide range equation of state for fluid hydrogen within density functional theory}
\author{Cong Wang}
\affiliation{Institute of Applied Physics and Computational
Mathematics, P.O. Box 8009, Beijing 100088, People's Republic of
China} \affiliation{Center for Applied Physics and Technology,
Peking University, Beijing 100871, People's Republic of China}
\author{Ping Zhang}
\thanks{Corresponding author: zhang\underline{
}ping@iapcm.ac.cn}
\affiliation{Institute of Applied Physics
and Computational Mathematics, P.O. Box 8009, Beijing 100088,
People's Republic of China} \affiliation{Center for Applied
Physics and Technology, Peking University, Beijing 100871,
People's Republic of China}


\begin{abstract}
Wide range equation of state (EOS) for liquid hydrogen is
ultimately built by combining two kinds of density functional
theory (DFT) molecular dynamics simulations, namely,
first-principles molecular dynamics simulations and orbital-free
molecular dynamics simulations. Specially, the present introduction
of short cutoff radius pseudopotentials enables the
hydrogen EOS to be available in the range $9.82\times10^{-4}$ to $1.347\times10^{3}$ g/cm$^{3}$ and up to $5\times10^{7}$ K.
By comprehensively comparing
with various attainable experimental and theoretical data, we
derive the conclusion that our DFT-EOS can be readily
and reliably conducted to hydrodynamic simulations of the inertial confinement
fusion.

\end{abstract}
\maketitle

\setcounter{MaxMatrixCols}{10} %TCIDATA{OutputFilter=LATEX.DLL}
%TCIDATA{Version=5.00.0.2552} %TCIDATA{<META NAME="SaveForMode"
%CONTENT="1">} %TCIDATA{Created=Thursday, January 17, 2008 16:18:47}
%TCIDATA{LastRevised=Thursday, July 17, 2008 19:00:53} %TCIDATA{<META
%NAME="GraphicsSave" CONTENT="32">} %TCIDATA{<META NAME="DocumentShell"
%CONTENT="Standard LaTeX\Blank - StandardLaTeX Article">}
%TCIDATA{Language=American English} %TCIDATA{CSTFile=40 LaTeX article.cst}
%TCIDATA{ComputeDefs= $A_{\sigma ,\Gamma =}$ $A_{\sigma ,\Gamma }=$ }

\setcounter{MaxMatrixCols}{10}

%\input{tcilatex}
%\affiliation{T03}

\bigskip

\section{Introduction}
Wide range equation of state (EOS) for hydrogen or its
isotopes is of crucial interest for inertial confinement fusion
(ICF) and astrophysics \cite{Atzeni2004,Guillot1999a,Guillot1999b}.
In the traditional central-hot-spot ignition designs of ICF, a
deuterium-tritium (D-T) capsule is assumed to be imploded to high
density either directly by high power laser pulses
\cite{McCrory2008} or indirectly by X rays generated in the hohlraum
\cite{Lindl1995}. Due to the fact that the compressibility of the
capsule is determined by EOS, high precision EOS of the D-T fuel is
essential for hydrodynamic simulations and ignition facility
designs. In astrophysics, the giant Jovian planets, such as Jupiter,
Saturn, Uranus, and Neptune, are composed primarily of hydrogen and
helium. The knowledge on the size and mass distribution of giant Jovian planets are
sensitive to the EOS of hydrogen in a wide range
\cite{Hubbard1980,Stevenson1982}.

The EOS of hydrogen has been probed through gas gun
\cite{Nellis2006}, converging explosive \cite{Boriskov2005},
magnetically driven flyer \cite{Knudson2004}, and high power
laser-driven experiments \cite{Collins1998,Boehly2004,Hicks2009},
where a pressure-temperature thermodynamical domain with amplitudes of megabar (Mbar) and electron volt (eV) has been reached. Theoretical
approximations, such as classical molecular dynamics based on
interatomic potentials \cite{Belonoshko2005}, linear mixing method
\cite{Ross1998}, fluid variational theory (FVT) \cite{Juranek2000},
path integral Monte Carlo (PIMC) \cite{Magro1996,Militzer2000}, and
quantum molecular dynamics (QMD) \cite{Bezkrovniy2004,Lenosky2000},
have already been employed to study high-pressure behaviors of
hydrogen and its isotopes. Although it has been pursued over
decades, there are still some fundamental issues worth to be
noticed. For instance, the first-order phase transition from
molecular to atomic fluid transition is being under intense
discussion, and plasma phase transition (PPT) characterized by
electronic ionization still needs to be clarified \cite{Nellis2006}.
High precision, wide range EOS for hydrogen are of particularly
importance for hydrodynamic simulations in ICF, especially at
densities from $\sim 10^{-3}$ to $10^{3}$ g/cm$^{3}$ and
temperatures up to 10$^{7}$ K, or even higher. Currently, SESAME-EOS
table \cite{Kerley1972,Kerley2003} for hydrogen describes chemical
species, such as, H$_{2}$ molecules, H atoms, and free protons and
electrons based on chemical models
\cite{Saumon1992,Ross1998,Rogers2001,Juranek2002}, which are only
expected to work well in the weak coupling limit. Recently, a new
EOS table based on PIMC simulations has been built \cite{{Hu2011}},
however, PIMC results are not consistent with experiments at
pressures below 50 GPa along the Hugoniot curve \cite{Magro1996}. As a
consequence, wide range EOS for hydrogen is highly recommended to
be constructed from other promising ways for comparisons and applications.

In the present work, a combined density functional theory (DFT) method of first-principles molecular
dynamics (FPMD) and orbital-free molecular dynamics (OFMD) has been used
to construct wide range EOS for fluid hydrogen with a temperature
range $10^{3}\sim5\times10^{7}$ K and density range $9.82\times10^{-4}\sim1.347\times10^{3}$ g/cm$^{3}$. In FPMD the
electrons are treated quantum mechanically through
finite-temperature DFT (FTDFT) with the only approximation of
exchange-correlation functional. Due to the Fermi-Dirac distribution
of the electronic states, at extremely high temperatures a huge number of
occupational bands have to be introduced, and FPMD simulations are
then restricted. As a consequence, OFMD simulations, where the electronic kinetic
energy is expressed as a functional of the local
electronic density and possibly of its gradient, have been adopted
to avoid the limitation. The rest of this paper is organized as
follows. Section II describes the computational methods with respect
to FPMD and OFMD. In section III, we discuss the EOS in detail, and
finally we get our conclusions in section IV.

\section{Computational Method}
In this section, we briefly describe the basic formalism employed
to explore thermodynamic properties of fluid hydrogen. That is,
two basic quantum-mechanical DFT approaches, one
based on Kohn-Sham (KS) formula and the other based on orbital-free
method. Then the simulation parameters are presented in
detail.

\subsection{First-principles molecular dynamics}
Our FPMD simulations for fluid hydrogen have been performed by using
ABINIT code \cite{abinitcode}. In these simulations, the
electrons are fully quantum mechanically treated by employing a
plane-wave FT-DFT description, where the electronic state occupations follow
the Fermi-Dirac distribution. The ions move classically according
to the forces from the electron density and the ion-ion repulsion.
We employed the NVT (canonical) ensemble,
where the number of particles $N$ and the volume are fixed
\cite{Haile1997}. The system was assumed to be in local
thermodynamic equilibrium with the electron and ion temperatures
being equal ($T_{e}=T_{i}$). In these calculations, the electronic
temperature was been kept constant according to the Fermi-Dirac
distribution, while the ionic temperature was controlled by the No\'{s}e
thermostat.

At each step during MD simulations, a set of electronic state
functions $\left\{  \Psi_{i,k}(r,t)\right\}  $ for each $\mathbf{k}$-point were determined
within KS construction by
\begin{equation}\label{equation_eigenfunction}
    H_{KS}\Psi_{i,k}(r,t)=\epsilon_{i,k}\Psi_{i,k}(r,t)
\end{equation}
with
\begin{equation}\label{equation_hamiltonian}
    H_{KS}=-\frac{1}{2}\nabla^{2}+V_{ext}+\int\frac{n(r')}{|r-r'|}dr'+v_{xc}(r),
\end{equation}
in which the four terms respectively represent the kinetic
contribution, the electron-ion interaction, the Hartree
contribution, and the exchange-correlation functional. The electronic density was obtained by
\begin{equation}\label{equation_electronicdensity}
    n(r)=\sum_{i,k}f_{i,k}|\Psi_{i,k}(r,t)|^{2}.
\end{equation}
Then by applying
the velocity Verlet algorithm, based on the force from
interactions between ions and electrons, a new set of positions
and velocities were obtained for ions.

\subsection{Orbital-free molecular dynamics}
OFMD simulations
\cite{Lambert2006,Mazevet2007,Lambert2008}, where
the kinetic energy of the electrons is treated semiclassically,
have also been used to investigate the wide range EOS for fluid hydrogen
under extreme conditions. The orbital-free electronic free energy
can be expressed as
\begin{equation}\label{equation_OFenergy}
\begin{split}
    F_{e}(n)=&\frac{1}{\beta}\int dr\{n(r)\Phi(n)-\frac{2\sqrt{2}}{3\pi^{2}\beta^{3/2}}I_{3/2}[\Phi(n)]\}+\int drh(n)\frac{|\bigtriangledown n|^{2}}{n}\\
    &+F_{xc}[n]+\frac{1}{2}\int\int drdr'\frac{n(r)n(r')}{|r-r'|}+\sum_{\ell=1}^{N_{\alpha}}Z_{\ell}\int dr\frac{n(r)}{r-R_{\ell}}\\
    &-\mu\int dr[n(r)-\sum_{\ell=1}^{N_{\alpha}}Z_{\ell}],
\end{split}
\end{equation}
where $I_{\nu}$ is the Fermi integral of order $\nu$, and the screened potential $\Phi$ is related
to the electronic density by
\begin{equation}\label{equation_conservation}
    \sum_{\ell=1}^{N_{\alpha}}Z_{\ell}=\frac{\sqrt{2}}{\pi^{2}\beta^{3/2}}\int drI_{1/2}(\Phi[n]).
\end{equation}
The first integral in Eq. (\ref{equation_OFenergy}), which depends only on the local electronic density
in the true spirit of the Hohenberg-Kohn theorem, is the well-known finite-temperature Thomas-Fermi expression \cite{Brack2003}.
The second term in Eq. (\ref{equation_OFenergy}) denotes the von Werzs\"{a}cker correction. In the present simulations we have omitted this gradient term and
worked in a Thomas-Fermi-Dirac form using the formula proposed by Perrot \cite{Perrot1979} to deal with the kinetic-entropic part. The orbital-free
procedure treats all electrons on an equal footing, albeit
approximately, with no distinction between bound and ionized
electrons. Except for that, the OFMD simulation procedure is similar to that of FPMD.

\subsection{Simulation details}

\begin{figure}[!ht]
\includegraphics[height=6.0cm]{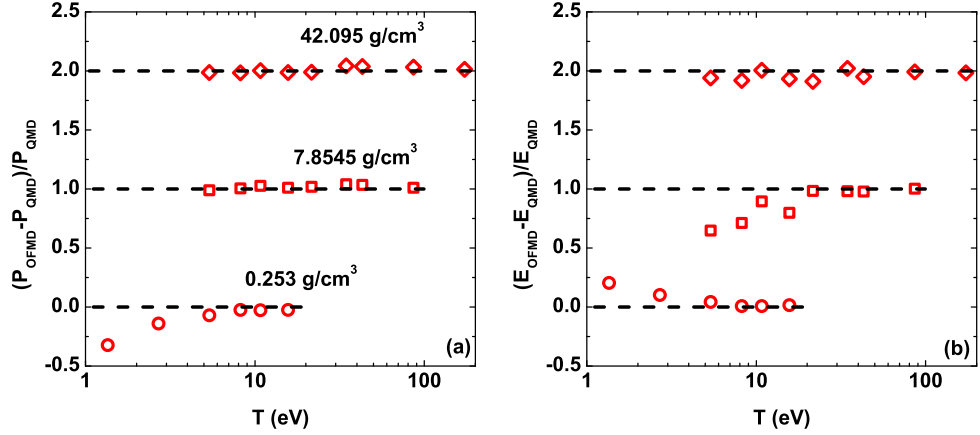}
\caption{(Color online) Pressure and internal energy
differences between QMD and OFMD methods as functions of
temperature at densities of 0.253 g/cm$^{3}$
(red open circles), 7.8545 g/cm$^{3}$ (red open squares), and 42.095
g/cm$^{3}$ (red open diamonds). QMD results have been plotted as
the black dashed line. Each curve corresponds to an isochore. Each
curve has been shifted by 1.0 from the previous one for
clarity.}\label{fig_qmd_ofmd_compare}
\end{figure}

Using the above-mentioned DFT formalisms (namely, FPMD and OFMD), we aim to build a wide range DFT-EOS table of data points for liquid hydrogen with the density ranging from
$9.82\times10^{-4}$ to $1.347\times10^{3}$ g/cm$^{3}$  and
temperature from $10^{3}$ to $5\times10^{7}$ K. Generally,
the Coulomb liquids can be characterized by two non-dimensional parameters.
That is, the ionic coupling parameter and electronic degenerate
parameter. For liquid hydrogen, the former one is commonly defined as
$\Gamma_{ii}=1/(k_{B}Ta)$, which presents the ratio of the mean
electrostatic potential energy and the mean kinetic energy of the
ions. The degeneracy parameter
$\theta=T/T_{F}$ is the ratio of the temperature to the Fermi
temperature $T_{F}=(3/\pi^{2}n_{e})^{2/3}/3$. Within the FPMD formalism,
the electronic states are occupied according to
the Fermi-Dirac distribution. Thus, our FPMD simulations have been restricted to
temperatures lower than $T_{F}$ ($\theta<1$) at
$\rho > 0.5$ g/cm$^{3}$. For lower densities, our FPMD simulations have been performed
up to a temperature of 15.682 eV. To overcome the computational
cost limit, OFMD was used in the same simulated conditions (density
and temperature) as those in FPMD, and explored to extend to higher
temperatures. The results indicate that both of the pressure and
internal energy difference are better than 2\% between QMD and
OFMD simulations as $\theta\sim1$ (see Fig.
\ref{fig_qmd_ofmd_compare}).

\begin{figure}[!ht]
\includegraphics[height=6.0cm]{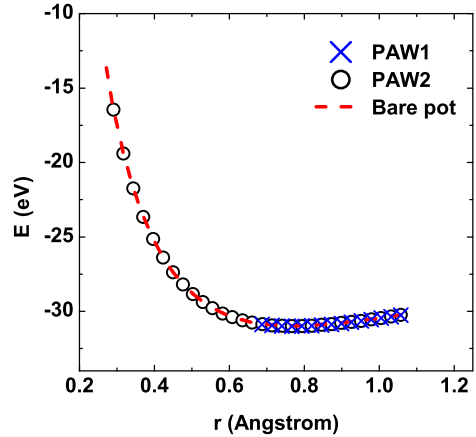}
\caption{(Color online) Calculated energy as a function of
interatomic distance. Results are obtained from PAW potentials
with a cutoff radius of 0.8 a.u. (PAW1), 0.1 a.u. (PAW2), and
Columbic potential (Bare pot), as labelled in the
figure.}\label{fig_pot}
\end{figure}

In general, FPMD and OFMD simulations based on DFT have introduced
pseudopotentials to reduce the computational cost and ensure the
accuracy at moderate densities. However, the pseudo-core
approximation fails at high densities, where the interatomic
distance is comparable with or smaller than the cutoff radius of
the pseudopotential, due to pressure-induced delocalization of the core
electrons and the overlapping of the pseudization spheres. In order to
avoid the limitations introduced by pseudopotential approximation,
a Columbic pseudopotential with a cutoff radius of 0.001 a.u. has
been built \cite{Wang2011,Wang2012}. As the energy
dependence is better than 1\% between projector augmented wave
(PAW) potentials and Columbic potential (see Fig. \ref{fig_pot}),
we explore the EOS of hydrogen into high density ($\sim10^{3}$
g/cm$^{3}$) by using a short cutoff radius Columbic potential.

We have considered a total number of 8 $\sim$ 512 atoms
(corresponding to expanded and ultra dense regimes) in a series of
volume-fixed supercells, which are repeated periodically
throughout the space. Only $\Gamma$ point is used to sample the
Brillouin zone in molecular dynamic simulations, because the
selection of higher number of $\mathbf{k}$ points modifies
the EOS within 3\%. Each system was assumed to be in local
thermodynamic equilibrium with the electron and ion temperatures
being equal ($T_{i}=T_{e}$). In order to balance the
pseudopotential approximation in the high density regime and the
computational cost, two potentials have been adopted in both FPMD and OFMD simulations.
That is, the PAW (with $R_{c}=0.1$
a.u.) pseudopotential ($\rho <$ 30 g/cm$^{3}$) and short cutoff
radius Coulombic potential ($\rho >$ 20 g/cm$^{3}$), where the
plane wave cutoff energy is set to 200 Ha, respectively. The
exchange-correlation functional is determined by local density
approximation (LDA) with Teter-Pade parametrization
\cite{Goedecker1996}, and the temperature dependence of
exchange-correlation functional, which is convinced to be as small
as negligible, is not taken into account. N$_{step}$ = 6000 has
been used in the molecular dynamic simulations, and the time steps
are selected with considering different density and temperature
\cite{timestep}. The EOS are obtained as running average of
the last 1000 steps of molecular dynamic simulations.
Additionally, in FPMD simulations, sufficient electronic states
have been adopted to secure the occupational number below
$10^{-6}$.

\section{Results And Discussion}

A wide range DFT-EOS (listed in Table I) has been constructed by data
obtained from FPMD (for $\theta < 1$) and OFMD ($\theta > 1$)
simulations. Results are compared with previous theoretical and
experimental ones in this section.
\begin{longtable}{r c c}
\caption{DFT-EOS table with pressure (GPa) and internal energy (eV/atom) for hydrogen.}\\
\hline \multicolumn{1}{c}{Temperature(eV)} &
\multicolumn{1}{c}{Pressure(GPa)} & \multicolumn{1}{c}{Internal
energy (eV/atom)} \\ \hline
\endfirsthead
\multicolumn{3}{c}%
{{\tablename\ \thetable{} -- continued from previous page}} \\
\hline\hline \multicolumn{1}{c}{Temperature(eV)} &
\multicolumn{1}{c}{Pressure(GPa)} & \multicolumn{1}{c}{Internal
energy (eV/atom)} \\ \hline
\endhead
\hline \multicolumn{3}{r}{{to be continued on next page}} \\
\hline
\endfoot
\hline \hline
\endlastfoot
\hline &$\rho=9.8200\times10^{-4}$g/cm$^{3}$ &\\
1.348   &   0.138   &   6.290   \\
2.695   &   0.351   &   13.021  \\
5.391   &   0.864   &   26.066  \\
8.215   &   1.393   &   35.787  \\
10.781  &   1.870   &   46.596  \\
15.682  &   2.698   &   62.313  \\
21.563  &   3.862   &   80.142  \\
34.500  &   6.316   &   119.051     \\
43.125  &   7.950   &   144.951     \\
86.250  &   16.114  &   274.346     \\
172.500     &   32.441  &   533.100     \\
345.000     &   65.097  &   1050.593    \\
690.000     &   130.407     &   2085.570    \\
1293.750    &   244.701     &   3896.774    \\
4312.500    &   816.153     &   12952.800   \\
\\
&$\rho=1.5590\times10^{-3}$g/cm$^{3}$ &\\
1.348   &   0.212   &   6.113   \\
2.695   &   0.542   &   12.339  \\
5.391   &   1.367   &   25.339  \\
8.215   &   2.239   &   35.584  \\
10.781  &   3.084   &   45.612  \\
15.682  &   4.431   &   61.651  \\
21.563  &   6.107   &   79.572  \\
34.500  &   10.004  &   118.522     \\
43.125  &   12.598  &   144.430     \\
86.250  &   25.561  &   273.838     \\
172.500     &   51.483  &   532.598     \\
345.000     &   103.324     &   1050.092    \\
690.000     &   207.002     &   2085.071    \\
1293.750    &   388.453     &   3896.275    \\
4312.500    &   1295.667    &   12952.301   \\
\\
&$\rho=2.6940\times10^{-3}$g/cm$^{3}$ &\\
1.348   &   0.359   &   5.927   \\
2.695   &   0.908   &   11.683  \\
5.391   &   2.298   &   24.074  \\
8.215   &   3.827   &   34.673  \\
10.781  &   5.188   &   44.264  \\
15.682  &   7.739   &   60.708  \\
21.563  &   10.493  &   78.773  \\
34.500  &   17.231  &   117.794     \\
43.125  &   21.715  &   143.715     \\
86.250  &   44.118  &   273.146     \\
172.500     &   88.912  &   531.914     \\
345.000     &   178.498     &   1049.412    \\
690.000     &   357.665     &   2084.391    \\
1293.750    &   671.210     &   3895.596    \\
4312.500    &   2238.934    &   12951.622   \\
\\
&$\rho=5.2620\times10^{-3}$g/cm$^{3}$ &\\
1.348   &   0.676   &   5.732   \\
2.695   &   1.710   &   10.860  \\
5.391   &   4.316   &   22.440  \\
8.215   &   7.278   &   33.087  \\
10.781  &   9.943   &   41.650  \\
15.682  &   14.981  &   59.245  \\
21.563  &   20.322  &   77.561  \\
34.500  &   33.495  &   116.710     \\
43.125  &   42.257  &   142.658     \\
86.250  &   86.019  &   272.126     \\
172.500     &   173.515     &   530.908     \\
345.000     &   348.498     &   1048.412    \\
690.000     &   698.457     &   2083.394    \\
1293.750    &   1310.879    &   3894.600    \\
4312.500    &   4373.030    &   12950.626   \\
\\
&$\rho=1.2473\times10^{-2}$g/cm$^{3}$ &\\
1.348   &   1.463   &   5.066   \\
2.695   &   3.884   &   9.320   \\
5.391   &   9.334   &   18.976  \\
8.215   &   16.559  &   30.688  \\
10.781  &   22.833  &   39.750  \\
15.682  &   34.787  &   56.895  \\
21.563  &   49.460  &   75.601  \\
34.500  &   80.720  &   115.000     \\
43.125  &   101.500     &   141.001     \\
86.250  &   204.800     &   270.546     \\
172.500     &   410.900     &   529.354     \\
345.000     &   823.100     &   1046.868    \\
690.000     &   1655.046    &   2081.854    \\
1293.750    &   3106.100    &   3893.061    \\
4312.500    &   10365.235   &   12949.087   \\
\\
&$\rho=2.1553\times10^{-2}$g/cm$^{3}$ &\\
1.348   &   2.472   &   4.493   \\
2.695   &   6.411   &   8.595   \\
5.391   &   16.101  &   18.019  \\
8.215   &   27.862  &   29.063  \\
10.781  &   38.592  &   37.668  \\
15.682  &   59.760  &   55.062  \\
21.563  &   84.600  &   73.978  \\
34.500  &   138.700     &   113.586     \\
43.125  &   174.600     &   139.635     \\
86.250  &   353.100     &   269.251     \\
172.500     &   709.600     &   528.081     \\
345.000     &   1425.665    &   1045.603    \\
690.000     &   2859.130    &   2080.591    \\
1293.750    &   5367.606    &   3891.798    \\
4312.500    &   17910.085   &   12947.823   \\
\\
&$\rho=4.2095\times10^{-2}$g/cm$^{3}$ &\\
1.348   &   4.264   &   4.712   \\
2.695   &   12.140  &   8.083   \\
5.391   &   30.746  &   16.663  \\
8.215   &   52.677  &   27.011  \\
10.781  &   73.257  &   35.436  \\
15.682  &   112.734     &   52.641  \\
21.563  &   162.600     &   71.692  \\
34.500  &   268.500     &   111.558     \\
43.125  &   339.100     &   137.679     \\
86.250  &   682.388     &   267.396     \\
172.500     &   1382.377    &   526.260     \\
345.000     &   2782.234    &   1043.791    \\
690.000     &   5581.767    &   2078.781    \\
1293.750    &   10481.126   &   3889.987    \\
4312.500    &   34977.421   &   12946.009   \\
\\
&$\rho=5.0\times10^{-2}$g/cm$^{3}$ &\\
1.348   &   5.218   &   4.035   \\
2.695   &   14.462  &   7.844   \\
5.391   &   33.584  &   17.101  \\
8.215   &   62.068  &   26.480  \\
10.781  &   86.394  &   34.845  \\
15.682  &   133.412     &   50.345  \\
21.563  &   192.000     &   71.017  \\
34.500  &   318.500     &   110.942     \\
43.125  &   401.800     &   137.082     \\
86.250  &   809.780     &   266.829     \\
172.500     &   1641.220    &   525.702     \\
345.000     &   3303.920    &   1043.235    \\
690.000     &   6629.240    &   2078.225    \\
1293.750    &   12448.600   &   3889.430    \\
4312.500    &   41545.000   &   12945.451   \\
\\
&$\rho=9.9781\times10^{-2}$g/cm$^{3}$ &\\
0.259   &   2.694   &   0.854   \\
0.431   &   3.948   &   1.153   \\
0.518   &   4.584   &   1.320   \\
0.690   &   5.537   &   1.717   \\
0.863   &   7.924   &   1.836   \\
1.348   &   12.306  &   3.131   \\
2.695   &   28.755  &   7.031   \\
5.391   &   71.782  &   14.766  \\
8.215   &   121.424     &   23.631  \\
10.781  &   167.717     &   32.418  \\
15.682  &   260.606     &   47.720  \\
21.563  &   375.300     &   67.930  \\
34.500  &   627.300     &   107.984     \\
43.125  &   794.300     &   134.196     \\
86.250  &   1609.411    &   264.053     \\
172.500     &   3268.481    &   522.956     \\
345.000     &   6586.363    &   1040.493    \\
690.000     &   13222.325   &   2075.477    \\
1293.750    &   24835.110   &   3886.676    \\
4312.500    &   82900.032   &   12942.686   \\
\\
&$\rho=1.5328\times10^{-1}$g/cm$^{3}$ &\\
0.259   &   4.773   &   0.752   \\
0.431   &   6.561   &   1.122   \\
0.518   &   7.336   &   1.311   \\
0.690   &   9.403   &   1.733   \\
0.863   &   10.704  &   2.281   \\
1.348   &   18.789  &   3.426   \\
2.695   &   45.908  &   6.683   \\
5.391   &   110.191     &   14.078  \\
8.215   &   185.120     &   22.566  \\
10.781  &   253.965     &   30.910  \\
15.682  &   395.081     &   45.969  \\
21.563  &   569.300     &   65.689  \\
34.500  &   955.800     &   105.708     \\
43.125  &   1212.000    &   131.937     \\
86.250  &   2489.000    &   261.841     \\
172.500     &   5013.911    &   520.751     \\
345.000     &   10110.722   &   1038.280    \\
690.000     &   20305.244   &   2073.253    \\
1293.750    &   38145.083   &   3884.443    \\
4312.500    &   127346.277  &   12940.439   \\
\\
&$\rho=1.9448\times10^{-1}$g/cm$^{3}$ &\\
0.259   &   7.194   &   0.779   \\
0.345   &   8.599   &   0.958   \\
0.431   &   9.692   &   1.133   \\
0.518   &   10.432  &   1.363   \\
0.690   &   13.249  &   1.753   \\
0.863   &   15.283  &   2.267   \\
1.348   &   26.465  &   3.124   \\
2.695   &   60.048  &   6.458   \\
5.391   &   135.419     &   14.176  \\
8.215   &   235.860     &   21.950  \\
10.781  &   320.826     &   30.080  \\
15.682  &   499.007     &   44.982  \\
21.563  &   720.600     &   64.364  \\
34.500  &   1209.000    &   104.306     \\
43.125  &   1537.000     &   130.534     \\
86.250  &   3158.000    &   260.445     \\
172.500     &   6396.000    &   519.350     \\
345.000     &   12849.073   &   1036.869    \\
690.000     &   25809.147   &   2071.833    \\
1293.750    &   48491.275   &   3883.014    \\
4312.500    &   161900.916  &   12938.999   \\
\\
&$\rho=2.5301\times10^{-1}$g/cm$^{3}$ &\\
0.259   &   11.871  &   0.798   \\
0.345   &   13.700  &   0.975   \\
0.431   &   15.444  &   1.162   \\
0.518   &   16.105  &   1.451   \\
0.690   &   18.874  &   1.889   \\
0.863   &   22.934  &   2.267   \\
1.348   &   36.898  &   3.296   \\
2.695   &   82.165  &   6.293   \\
5.391   &   186.413     &   13.224  \\
8.215   &   298.231     &   21.654  \\
10.781  &   414.411     &   29.201  \\
15.682  &   643.852     &   43.913  \\
21.563  &   924.300     &   62.908  \\
34.500  &   1562.000    &   102.714     \\
43.125  &   1984.000    &   128.927     \\
86.250  &   4094.000    &   258.819     \\
172.500     &   8293.000    &   517.709     \\
345.000     &   16672.990   &   1035.209    \\
690.000     &   33498.380   &   2070.157    \\
1293.750    &   62944.837   &   3881.326    \\
4312.500    &   210176.125  &   12937.294   \\
\\
&$\rho=3.3676\times10^{-1}$g/cm$^{3}$ &\\
0.259   &   22.170  &   0.881   \\
0.345   &   24.459  &   1.064   \\
0.431   &   24.239  &   1.349   \\
0.518   &   26.790  &   1.617   \\
0.690   &   31.156  &   1.949   \\
0.863   &   37.359  &   2.291   \\
1.348   &   57.070  &   3.257   \\
2.695   &   117.535     &   6.153   \\
5.391   &   255.253     &   12.911  \\
8.215   &   414.266     &   20.725  \\
10.781  &   566.886     &   28.198  \\
15.682  &   852.768     &   42.767  \\
21.563  &   1220.000    &   61.324  \\
34.500  &   2066.000    &   100.926     \\
43.125  &   2627.000    &   127.094     \\
86.250  &   5430.000    &   256.931     \\
172.500     &   10984.218   &   515.784     \\
345.000     &   22178.636   &   1033.253    \\
690.000     &   44572.273   &   2068.174    \\
1293.750    &   83762.886   &   3879.325    \\
4312.500    &   279722.953  &   12935.268   \\
\\
&$\rho=4.1867\times10^{-1}$g/cm$^{3}$ &\\
0.086   &   28.457  &   0.666   \\
0.259   &   36.339  &   0.954   \\
0.431   &   36.579  &   1.480   \\
0.690   &   47.351  &   2.009   \\
0.863   &   55.779  &   2.342   \\
1.348   &   81.633  &   3.291   \\
2.695   &   156.889     &   6.117   \\
5.391   &   326.268     &   12.694  \\
8.215   &   522.548     &   20.335  \\
10.781  &   685.171     &   27.621  \\
15.682  &   1058.104    &   41.926  \\
21.563  &   1515.000    &   60.154  \\
34.500  &   2562.000    &   99.553  \\
43.125  &   3267.000    &   125.671     \\
86.250  &   6744.000    &   255.430     \\
172.500     &   13690.000   &   514.238     \\
345.000     &   27562.538   &   1031.671    \\
690.000     &   55403.077   &   2066.565    \\
1293.750    &   104128.019  &   3877.696    \\
4312.500    &   347756.729  &   12933.615   \\
\\
&$\rho=5.0\times10^{-1}$g/cm$^{3}$ &\\
0.086   &   43.727  &   0.739   \\
0.259   &   49.036  &   1.148   \\
0.431   &   53.691  &   1.583   \\
0.690   &   68.576  &   2.092   \\
0.863   &   79.054  &   2.419   \\
1.348   &   110.256     &   3.352   \\
2.695   &   203.690     &   6.226   \\
5.391   &   403.036     &   12.590  \\
8.215   &   634.218     &   20.070  \\
10.781  &   854.595     &   27.285  \\
15.682  &   1293.154    &   41.537  \\
21.563  &   1799.000    &   59.237  \\
34.500  &   3020.820    &   98.444  \\
43.125  &   3844.600    &   124.507     \\
86.250  &   7984.900    &   254.174     \\
172.500     &   16287.600   &   512.930     \\
345.000     &   32907.200   &   1030.326    \\
690.000     &   66155.400   &   2065.190    \\
1293.750    &   124347.000  &   3876.302    \\
4312.500    &   415320.000  &   12932.195   \\
\\
&$\rho=7.9825\times10^{-1}$g/cm$^{3}$ &\\
1.348   &   259.313     &   3.792   \\
2.695   &   404.291     &   6.492   \\
5.391   &   718.712     &   12.638  \\
8.215   &   1077.005    &   19.843  \\
10.781  &   1423.714    &   26.707  \\
15.682  &   2108.910    &   40.544  \\
21.563  &   2905.275    &   57.192  \\
34.500  &   4831.500    &   95.738  \\
43.125  &   6136.150    &   121.574     \\
86.250  &   12721.700   &   250.794     \\
172.500     &   25961.000   &   509.293     \\
345.000     &   52483.001   &   1026.515    \\
690.000     &   105556.001  &   2061.257    \\
1293.750    &   198451.502  &   3872.289    \\
4312.500    &   662975.008  &   12928.083   \\
\\
&$\rho=9.8181\times10^{-1}$g/cm$^{3}$ &\\
1.348   &   384.866     &   4.176   \\
2.695   &   560.854     &   6.845   \\
5.391   &   944.470     &   12.880  \\
8.215   &   1382.305    &   19.957  \\
10.781  &   1796.059    &   26.734  \\
15.682  &   2632.108    &   40.258  \\
21.563  &   3678.614    &   57.308  \\
34.500  &   5871.477    &   93.996  \\
43.125  &   7456.522    &   120.184     \\
86.250  &   15556.243   &   252.157     \\
172.500     &   31880.886   &   512.898     \\
345.000     &   64517.773   &   1030.922    \\
690.000     &   129799.546  &   2065.866    \\
1293.750    &   244056.648  &   3876.955    \\
4312.500    &   815392.161  &   12932.766   \\
\\
&$\rho$=1.2263 g/cm$^{3}$ &\\
1.348   &   591.578     &   4.774   \\
2.695   &   811.125     &   7.441   \\
5.391   &   1282.332    &   13.347  \\
8.215   &   1816.492    &   20.228  \\
10.781  &   2324.122    &   26.844  \\
15.682  &   3353.252    &   40.091  \\
21.563  &   4638.863    &   56.811  \\
34.500  &   7391.677    &   92.514  \\
43.125  &   9309.797    &   118.716     \\
86.250  &   19393.895   &   250.119     \\
172.500     &   39770.389   &   510.828     \\
345.000     &   80525.778   &   1028.729    \\
690.000     &   162054.556  &   2063.622    \\
1293.750    &   304757.918  &   3874.666    \\
4312.500    &   1018329.725     &   12930.412   \\
\\
&$\rho$=1.5591 g/cm$^{3}$ &\\
2.695   &   1217.205    &   8.329   \\
5.391   &   1802.722    &   14.093  \\
8.215   &   2471.084    &   20.813  \\
10.781  &   3107.909    &   27.280  \\
15.682  &   4394.780    &   40.431  \\
21.563  &   6008.367    &   56.715  \\
34.500  &   9711.442    &   94.365  \\
43.125  &   11868.865   &   117.384     \\
86.250  &   24627.130   &   248.026     \\
172.500     &   50504.260   &   508.563     \\
345.000     &   102305.520  &   1026.416    \\
690.000     &   205947.039  &   2061.235    \\
1293.750    &   387371.949  &   3872.216    \\
4312.500    &   1294596.497     &   12927.871   \\
\\
&$\rho$=2.0241 g/cm$^{3}$ &\\
2.695   &   1914.054    &   9.701   \\
5.391   &   2653.269    &   15.315  \\
8.215   &   3499.908    &   21.844  \\
10.781  &   4303.718    &   28.115  \\
15.682  &   5940.816    &   40.924  \\
21.563  &   8003.076    &   56.885  \\
34.500  &   12739.349   &   93.854  \\
43.125  &   15981.851   &   119.108     \\
86.250  &   31973.372   &   245.712     \\
172.500     &   65497.743   &   505.963     \\
345.000     &   132718.486  &   1023.755    \\
690.000     &   267252.973  &   2058.470    \\
1293.750    &   502770.574  &   3869.360    \\
4312.500    &   1680568.581     &   12924.885   \\
\\
&$\rho$=3.6956 g/cm$^{3}$ &\\
2.695   &   5561.156    &   15.030  \\
5.391   &   6851.582    &   20.465  \\
8.215   &   8282.303    &   26.503  \\
10.781  &   9683.735    &   32.433  \\
15.682  &   12471.118   &   44.273  \\
21.563  &   16109.020   &   59.548  \\
34.500  &   24440.344   &   94.995  \\
43.125  &   30151.959   &   119.149     \\
86.250  &   58805.086   &   239.239     \\
172.500     &   119458.718  &   500.123     \\
345.000     &   241975.435  &   1017.331    \\
690.000     &   487450.871  &   2051.576    \\
1293.750    &   917350.383  &   3862.109    \\
4312.500    &   3067567.944     &   12917.150   \\
\\
&$\rho$=5.0 g/cm$^{3}$ &\\
2.695   &   9549.150    &   19.264  \\
5.391   &   11289.594   &   24.705  \\
8.215   &   13149.412   &   30.532  \\
10.781  &   14947.198   &   36.149  \\
15.682  &   18635.588   &   47.707  \\
21.563  &   23350.696   &   62.524  \\
34.500  &   33750.403   &   94.172  \\
43.125  &   41345.121   &   119.872     \\
86.250  &   80760.602   &   244.172     \\
172.500     &   162974.601  &   496.972     \\
345.000     &   329461.530  &   1026.672    \\
690.000     &   662036.697  &   2067.672    \\
1293.750    &   1243990.000     &   3869.013    \\
4312.500    &   4152780.000     &   12918.432   \\
\\
&$\rho$=7.8545 g/cm$^{3}$ &\\
5.391   &   23919.083   &   32.364  \\
8.215   &   26366.144   &   35.213  \\
10.781  &   28417.191   &   40.201  \\
15.682  &   34567.155   &   50.461  \\
21.563  &   40887.110   &   63.836  \\
34.500  &   57821.094   &   96.066  \\
43.125  &   69405.461   &   118.905     \\
86.250  &   129424.016  &   240.005     \\
172.500     &   257431.789  &   492.021     \\
345.000     &   517526.963  &   968.205     \\
690.000     &   1040160.000     &   2052.492    \\
1293.750    &   1953890.000     &   3857.032    \\
4312.500    &   6522980.000     &   12881.855   \\
\\
&$\rho=1.2473\times10^{1}$ g/cm$^{3}$ &\\
5.391   &   52334.071   &   45.782  \\
8.215   &   56804.007   &   50.849  \\
10.781  &   60424.429   &   54.502  \\
15.682  &   68219.429   &   62.075  \\
21.563  &   77112.745   &   74.889  \\
34.500  &   103991.965  &   105.034     \\
43.125  &   121078.467  &   127.234     \\
86.250  &   213335.193  &   245.844     \\
172.500     &   412180.825  &   487.534     \\
345.000     &   823130.585  &   1015.634    \\
690.000     &   1652820.000     &   2028.342    \\
1293.750    &   3103200.000     &   3841.490    \\
4312.500    &   10356600.000    &   12898.888   \\
\\
&$\rho=2.1553\times10^{1}$ g/cm$^{3}$ &\\
5.391   &   132965.313  &   69.516  \\
8.215   &   140631.941  &   75.069  \\
10.781  &   147606.997  &   79.707  \\
15.682  &   161508.892  &   91.188  \\
21.563  &   176692.441  &   98.682  \\
34.500  &   214260.781  &   124.801     \\
43.125  &   244807.793  &   144.762     \\
86.250  &   397368.524  &   255.511     \\
172.500     &   733042.082  &   498.257     \\
345.000     &   1436410.000     &   1015.776    \\
690.000     &   2860880.000     &   2037.937    \\
1293.750    &   5365630.000     &   3849.903    \\
4312.500    &   17899300.000    &   12890.991   \\
\\
&$\rho=4.2095\times10^{1}$ g/cm$^{3}$ &\\
5.391   &   420619.812  &   120.275     \\
8.215   &   436741.892  &   121.507     \\
10.781  &   445038.475  &   125.749     \\
15.682  &   477946.889  &   134.127     \\
21.563  &   500374.227  &   144.655     \\
34.500  &   574053.275  &   169.598     \\
43.125  &   621244.176  &   187.496     \\
86.250  &   900873.051  &   288.390     \\
172.500     &   1516040.000     &   519.066     \\
345.000     &   2861940.000     &   996.990     \\
690.000     &   5626920.000     &   2054.507    \\
1293.750    &   10503500.000    &   3853.605    \\
4312.500    &   34970300.000    &   12898.971   \\
\\
&$\rho=8.4190\times10^{1}$ g/cm$^{3}$ &\\
5.391   &   1380830.000     &   197.188     \\
8.215   &   1405720.000     &   199.217     \\
10.781  &   1437320.000     &   206.732     \\
15.682  &   1491360.000     &   217.589     \\
21.563  &   1546100.000     &   223.268     \\
34.500  &   1657160.000     &   252.920     \\
43.125  &   1765120.000     &   269.112     \\
86.250  &   2250500.000     &   359.484     \\
172.500     &   3390870.000     &   572.109     \\
345.000     &   5962050.000     &   1047.419    \\
690.000     &   11406900.000    &   2049.414    \\
1293.750    &   21127400.000    &   3824.306    \\
4312.500    &   69971100.000    &   12895.733   \\
\\
&$\rho=1.6838\times10^{2}$ g/cm$^{3}$ &\\
5.391   &   4470890.000     &   348.905     \\
8.215   &   4531710.000     &   353.258     \\
10.781  &   4567730.000     &   357.249     \\
15.682  &   4706480.000     &   365.018     \\
21.563  &   4754720.000     &   374.545     \\
34.500  &   5038440.000     &   396.244     \\
43.125  &   5235950.000     &   411.271     \\
86.250  &   6139410.000     &   492.918     \\
172.500     &   8286640.000     &   684.142     \\
345.000     &   13024000.000    &   1128.948    \\
690.000     &   23505400.000    &   2101.594    \\
1293.750    &   42790700.000    &   3685.375    \\
4312.500    &   140264000.000   &   12902.503   \\
\\
&$\rho=3.3676\times10^{2}$ g/cm$^{3}$ &\\
5.391   &   14545500.000    &   588.613     \\
8.215   &   14665000.000    &   591.546     \\
10.781  &   14723000.000    &   599.508     \\
15.682  &   14995000.000    &   602.836     \\
21.563  &   15179700.000    &   616.630     \\
34.500  &   15789200.000    &   636.221     \\
43.125  &   16093600.000    &   650.506     \\
86.250  &   17859500.000    &   726.080     \\
172.500     &   21530900.000    &   897.457     \\
345.000     &   30504600.000    &   1301.652    \\
690.000     &   50560500.000    &   2224.865    \\
1293.750    &   87731800.000    &   3952.434    \\
4312.500    &   281715000.000   &   12945.568   \\
\\
&$\rho=6.7352\times10^{2}$ g/cm$^{3}$ &\\
5.391   &   47188700.000    &   985.291     \\
8.215   &   47374300.000    &   989.586     \\
10.781  &   47542300.000    &   993.478     \\
15.682  &   47866200.000    &   1000.964    \\
21.563  &   48258700.000    &   1010.034    \\
34.500  &   49138100.000    &   1030.355    \\
43.125  &   49735000.000    &   1044.162    \\
86.250  &   52836300.000    &   1115.850    \\
172.500     &   59610600.000    &   1272.441    \\
345.000     &   75258300.000    &   1634.074    \\
690.000     &   112078000.000   &   2485.051    \\
1293.750    &   183785000.000   &   4142.350    \\
4312.500    &   566346000.000   &   12930.233   \\
\\
&$\rho=1.3470\times10^{3}$ g/cm$^{3}$ &\\
5.391   &   152939000.000   &   1629.645    \\
8.215   &   153307000.000   &   1633.926    \\
10.781  &   153649000.000   &   1637.809    \\
15.682  &   152494000.000   &   1645.251    \\
21.563  &   155067000.000   &   1654.214    \\
34.500  &   156786000.000   &   1674.139    \\
43.125  &   157956000.000   &   1687.602    \\
86.250  &   163941000.000   &   1756.777    \\
172.500     &   176641000.000   &   1903.589    \\
345.000     &   204903000.000   &   2230.145    \\
690.000     &   271115000.000   &   2995.092    \\
1293.750    &   405010000.000   &   4541.812    \\
4312.500    &   1154030000.000  &   12958.110   \\
\end{longtable}

\subsection{Hugoniot curve}

\begin{figure}[!ht]
\includegraphics[height=6.0cm]{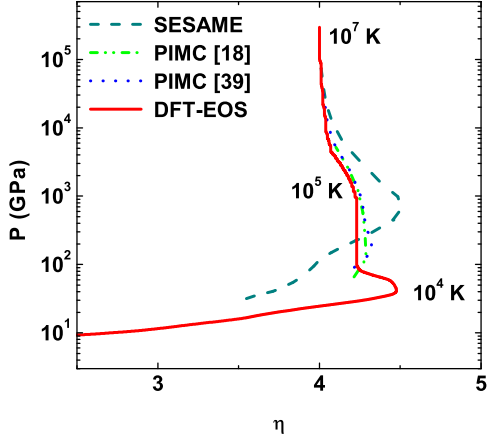}
\caption{(Color online) Present DFT-EOS for wide-range
Hugoniot curve (red line) of liquid hydrogen. Previous wide-range Hugoniot curves from PIMC simulations (by S. X. Hu \emph{et al.}
\cite{Hu2010} and Militzer \emph{et al.}
\cite{Militzer2000}) and SESAME \cite{Kerley1972} are
also shown for comparison.}\label{fig_hugoniot1}
\end{figure}

High precision EOS data are essential for understanding target
implosion process in ICF. We first examine the present DFT-EOS
theoretically through the Rankine-Hugoniot (RH) equations, which
follow from conservation of mass, momentum, and energy across the
front of the shock wave. The locus of points in ($E$, $P$, $V$)-space
described by RH equations satisfy

\begin{equation}\label{equation_hugoniot}
    E_{1}-E_{0}=\frac{1}{2}(P_{1}+P_{0})(V_{1}-V_{0}),
\end{equation}
\begin{equation}\label{equation_hugoniot_P}
    (P_{1}-P_{0})=\rho_{0}u_{s}u_{p},
\end{equation}
\begin{equation}\label{equation_hugoniot_V}
    V_{1}=V_{0}(1-u_{p}/u_{s}),
\end{equation}
where subscripts 0 and 1 represent the initial and shocked state,
and $E$, $P$, and $V$ denote internal energy, pressure, and volume,
respectively. $u_{p}$ is the particle velocity of the material behind the shock front and $u_{s}$ is the shock velocity. Along the Hugoniot curve of the liquid hydrogen, the
starting point with a density of 0.0855 g/cm$^{3}$ and a temperature of 23 K has been
selected, where the relative internal energy has been set to zero
and the pressure is considered as small as negligible. Smooth
functions have been adopted to fit DFT-EOS in the relative density
and temperature regime. Our DFT-based Hugoniot
curve with pressure up to $10^{5}$ GPa  and temperature up to $5\times10^{7}$ K is presented in Fig. \ref{fig_hugoniot1}.
Our simulation results indicate that the maximum shock compression ratio is 4.5 with a pressure around 40 GPa, at which the system is governed by gradual dissociation of molecules.
With the increase of
pressure, the compression ratio decreases and then reaches a
value of 4.23 below 950 GPa. This hardening behavior of the
Hugoniot can be attributed to the formation of mono-atomic fluid.
However, as the pressure exceeds $10^{3}$ GPa (temperature above 19 eV), the
compression ratio merges into 4.0, which indicate a full
ionization of the liquid hydrogen. On the other side, SESAME Hugoniot
\cite{Kerley1972} plotted in Fig. \ref{fig_hugoniot1} shows a maximum compression ratio around 4.5,
but the corresponding pressure is much too high with respect to our DFT-Hugoniot. For comparison, previous results from PIMC
simulations \cite{Militzer2000,Hu2010} are also plotted in
Fig. \ref{fig_hugoniot1}, which shows consistency with our DFT results at pressures beyond 50 GPa. However, PIMC simulations have failed to reproduce the experimental results below 50 GPa.

\begin{figure}[!ht]
\includegraphics[height=6.0cm]{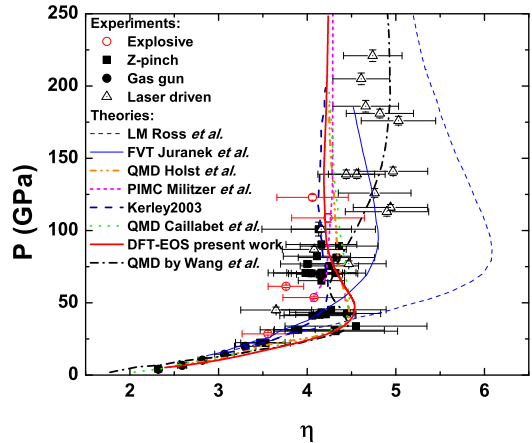}
\caption{(Color online) Hugoniot curve based on present DFT-EOS
for liquid hydrogen with (red solid line) or without (red dashed
line) considering ionic quantum zero-point energy. Previous experimental results
and theoretical predictions are shown for comparison. Experimental
data: gas gun by Nellis \emph{et al.} \cite{Nellis2006}
(solid circles), Z-pinch by Knudson \emph{et al.}
\cite{Knudson2004} (solid squares), explosives of Boriskov
\emph{et al.} \cite{Boriskov2005} (open circles), laser-driven
by Hicks \emph{et al.} \cite{Hicks2009} (up open triangles),
and Boehly \emph{et al.} \cite{Boehly2004} (down open
triangles). Theoretical data: QMD simulation results by Holst
\emph{et al.} (orange line) \cite{Holst2008}, Caillabet
\emph{et al.} (green line) \cite{Caillabet2011}, and Wang \emph{et al.}
(black dash-dotted line) \cite{Wang2010}, PIMC
results by Militzer \emph{et al.} (magenta line)
\cite{Militzer2000}, Kerley (royal line)
\cite{Kerley2003}, LM model by Ross (blue dashed line)
\cite{Ross1998}, and FVT (blue solid line)
\cite{Juranek2002}.}\label{fig_hugoniot2}
\end{figure}

Over the past ten years, the Hugoniot of hydrogen or deuterium has
been experimentally explored up to $\sim$200 GPa. The latest
set of data points were obtained by two-stage light gas gun
\cite{Nellis2006}, explosive-driven compression
\cite{Boriskov2005}, Z-pinch-driven compression
\cite{Knudson2004}, where the compression ratio $\eta$ shows
a maximum close to 4.3, or by laser-driven compression with the
Nova laser and the Omega laser (the EOS would possibly be
corrected by introducing the quartz standard), which suggest a
stiff behavior ($\eta_{max}\approx4.2$) below 100 GPa and become
softer ($\eta_{max}\approx4.5\sim5.5$) at higher pressures
\cite{Hicks2009,Boehly2004}. To clearly show the
comparison between the present DFT-EOS and those previous results,
Fig. (\ref{fig_hugoniot2}) plots the Hugoniot curve below 250 GPa.
The present Hugoniot curve from DFT-EOS with accounting for the ionic
quantum zero-point energy (ZPE) shows better accordance with experimental
data. At pressures below 100 GPa, both of the curves, as discussed
above, exhibit a maximum compression ratio of 4.5, which is
accordant with previous QMD results and experimental data obtained
by gas gun, converging explosives and magnetically driven flyer.
However, as pressures go beyond 100 GPa, $\eta\sim4.3$ indicates the
agreement with high power laser experiments with the quartz
standard. Predictions from various chemical models are also shown
for comparison, for instance, Kerley 2003 \cite{Kerley2003},
linear mixing model \cite{Ross1998}, and the fluid
variational theory (FVT) \cite{Juranek2002}. Those chemical
methods generally predict larger maximum compression ratio into
higher pressures, except for the Kerley 2003 EOS, which is in
better agreement with experiments.

\begin{figure}[!ht]
\includegraphics[height=6.0cm]{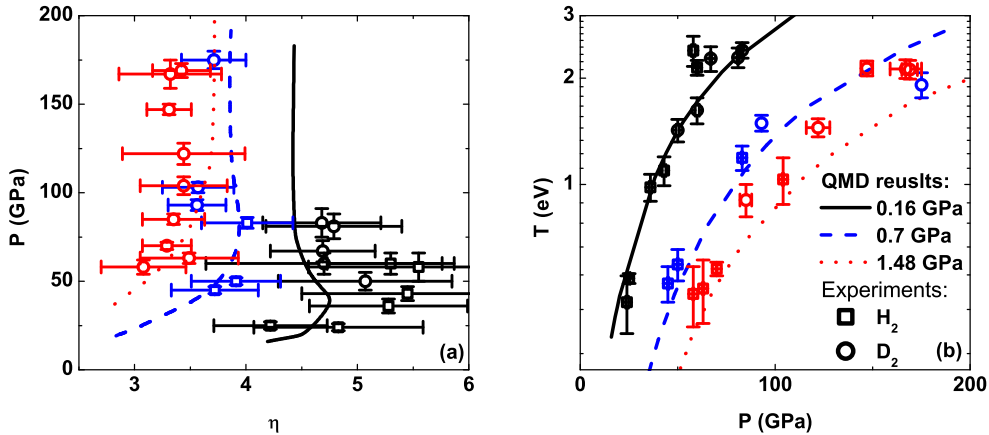}
\caption{(Color online) Pressure versus compression ratio (a) and
temperature versus pressure (b). For comparison with the most recent experimental data \cite{Loubeyre2012}, three initial pre-compressions of
H$_{2}$ (D$_{2}$) samples at 297 K have been studied: 0.16 GPa,
0.7 GPa, and 1.48 GPa.}\label{fig_hugoniot3}
\end{figure}

Recently, laser-driven shock compressions on H$_{2}$ or D$_{2}$
precompressed in diamond anvil cells from 0.16 to 1.6 GPa have been
proved to provide visible ways to generate shock Hugoniot data over
a significantly broader thermodynamical regime than previous
experiments \cite{Loubeyre2012}. These experimental data are highly valuable
for examining various theoretical models. In the
present work, we have shown the Hugoniot data for initial
pre-compressions of 0.16 GPa, 0.7 GPa, and 1.48 GPa in Fig.
\ref{fig_hugoniot3}. As a consequence, the data covers a density
range $\sim$ 2 times greater than previous investigations limited
to the principal Hugoniot alone. As shown in Fig.
\ref{fig_hugoniot3}, good agreement has been gained between the
present DFT-EOS pre-compressed Hugoniot data and laser-driven
experimental results. The maximum compression ratio along a given
Hugoniot has been observed to strongly depend on the initial
density. That is, with increasing initial density, the
compression ratio decreases.

\subsection{Molecular dissociations}

For molecular fluid in the warm dense regime, which consists of
atoms, molecules, nuclei, and electrons, the free energy can be
expressed as
\begin{equation}\label{equation_freeenergy}
    F(\rho,T)=F_{id}^{(i)}+F_{id}^{(e)}+F_{ex}^{(i-e)}+F_{dis}^{(mol)},
\end{equation}
where $F_{id}^{(i)}$ and $F_{id}^{(e)}$ are the ideal free energies
for ions and electrons, respectively, $F_{ex}^{(i-e)}$ is the excess
free energy, while $F_{dis}^{(mol)}$ denotes the contribution from
molecular dissociation with the following form \cite{Caillabet2011}:
\begin{equation}\label{equation_freeenergy_dis}
    F_{dis}^{(mol)}=Nk_{B}T  \left\{  \ln\alpha+\frac{1-\alpha}{2}\right\}.
\end{equation}
Here the dissociation ratio $\alpha$ is used as an adjustable function of density and temperature, with an assumed Fermi-function form
\begin{equation}\label{equation_alpha}
    \alpha(\rho,T)=\frac{1}{\exp[B(\rho)/T-C(\rho)T]+1},
\end{equation}
where $B(\rho)=\exp(B_{1}+B_{2}\rho)$ and
$C(\rho)=\exp(C_{1}+C_{2}\rho)$. Using the present DFT-EOS data,
we have determined the value $B_{1}=9.5517$, $B_{2}=-2.8277$,
$C_{1}=-8.2946$, and $C_{2}=0.4708$.

\begin{figure}[!ht]
\includegraphics[height=6.0cm]{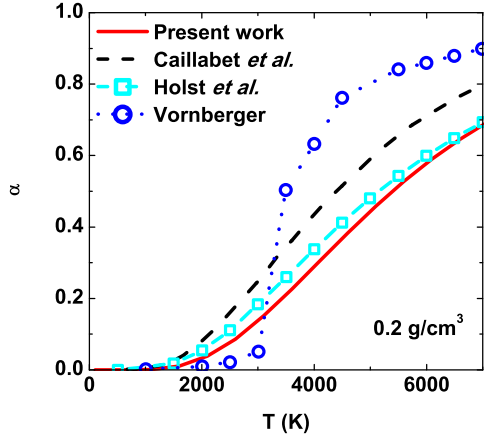}
\caption{(Color online) Calculated molecular dissociation fraction as a function of temperature at 0.2 g/cm$^{3}$.
For comparison, previous theoretical results \cite{Holst2008,Caillabet2011,Vorberger2007} have also been plotted.}\label{fig_dissociation}
\end{figure}

At temperatures below 10000 K, molecular dissociation governs the
first-order phase transition, which is important in determining the
nonmetal-to-metal transition. In this work, we have introduced a
Fermi formula to fit our DFT-EOS in warm dense region by using Eq.
(\ref{equation_alpha}), and the dissociation fraction has been
plotted in Fig. \ref{fig_dissociation}. Vorberger \emph{et al.}
\cite{Vorberger2007} have introduced a criteria to estimate the fraction
of molecular hydrogen by counting the number of atoms located within
a radius, which last for a time greater than ten vibrational
periods. Holst \emph{et al.} \cite{Holst2008} have used a coordination
number,
\begin{equation}\label{equation_coordination}
    K(r)=\frac{N-1}{V}\int_{0}^{r}4\pi r'^{2}g(r')dr',
\end{equation}
to determine $\alpha$, where $g(r)$ is the pair correlation
function to present the possibility of finding a particle from a
reference atom.  Results from those different assumptions are
compared with the present work in Fig. \ref{fig_dissociation}. They yield similar tendency with
temperature at the sampled density. The dissociation fraction
from Vorberger \emph{et al.} strongly depends on the definition
of the molecule in this region, and shows abrupt increase as
temperature increases. However, QMD method gives a smoother
behavior of $\alpha$ as indicated in Fig. \ref{fig_dissociation}.

\subsection{Comparison of DFT-EOS with previous theoretical results}

\begin{figure}[!ht]
\includegraphics[height=6.0cm]{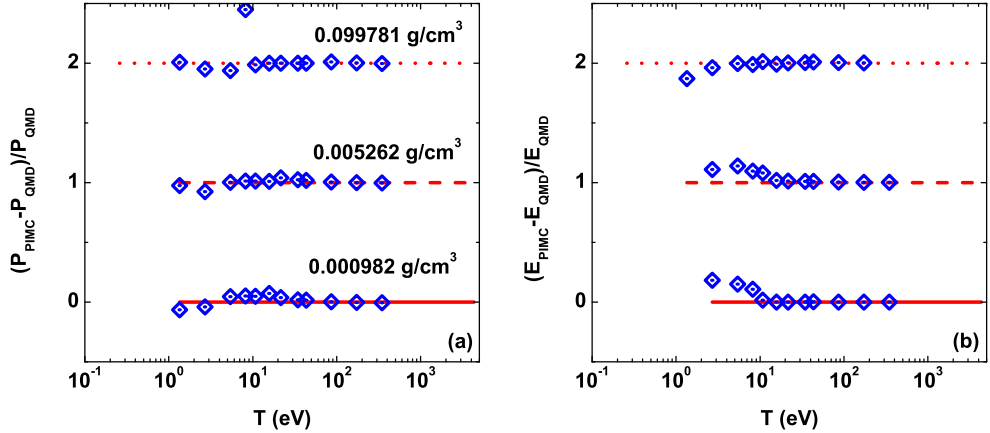}
\caption{(Color online) Pressure (a) and internal energy (b)
differences between QMD and PIMC \cite{Hu2011} methods as functions of temperature.  QMD results are plotted as the red
solid, red dashed, and red dotted lines at densities of 0.000982,
0.005262, and 0.099781 g/cm$^{3}$, respectively. The blue open
diamonds denote PIMC data. Each curve corresponds to an isochore. Each curve has been shifted by 1.0 from the previous one for
clarity.}\label{fig_expanded}
\end{figure}

\begin{figure}[!ht]
\includegraphics[height=6.0cm]{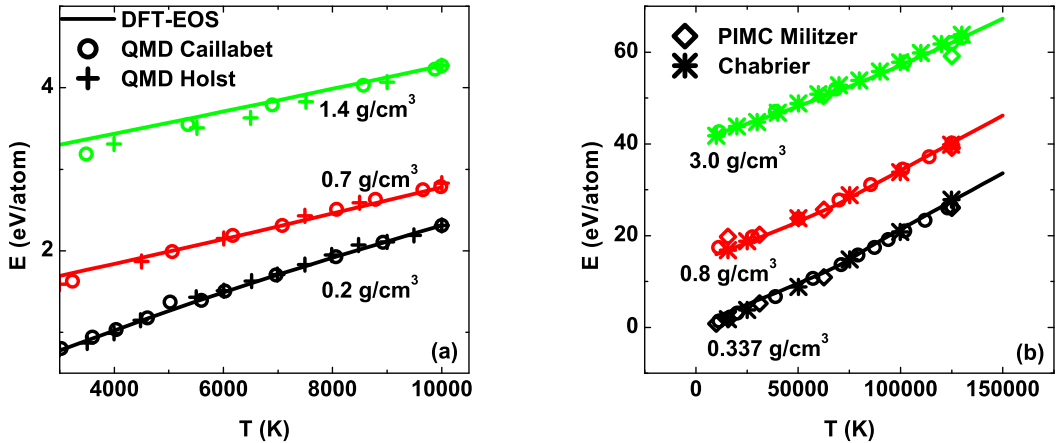}
\caption{(Color online) Internal energy of liquid hydrogen as a function of
temperature. Previous QMD
\cite{Holst2008,Caillabet2011}, PIMC
\cite{Militzer2000}, and Chabrier model
\cite{Chabrier1998} results are also shown for comparison.}\label{fig_E1}
\end{figure}

\begin{figure}[!ht]
\includegraphics[height=6.0cm]{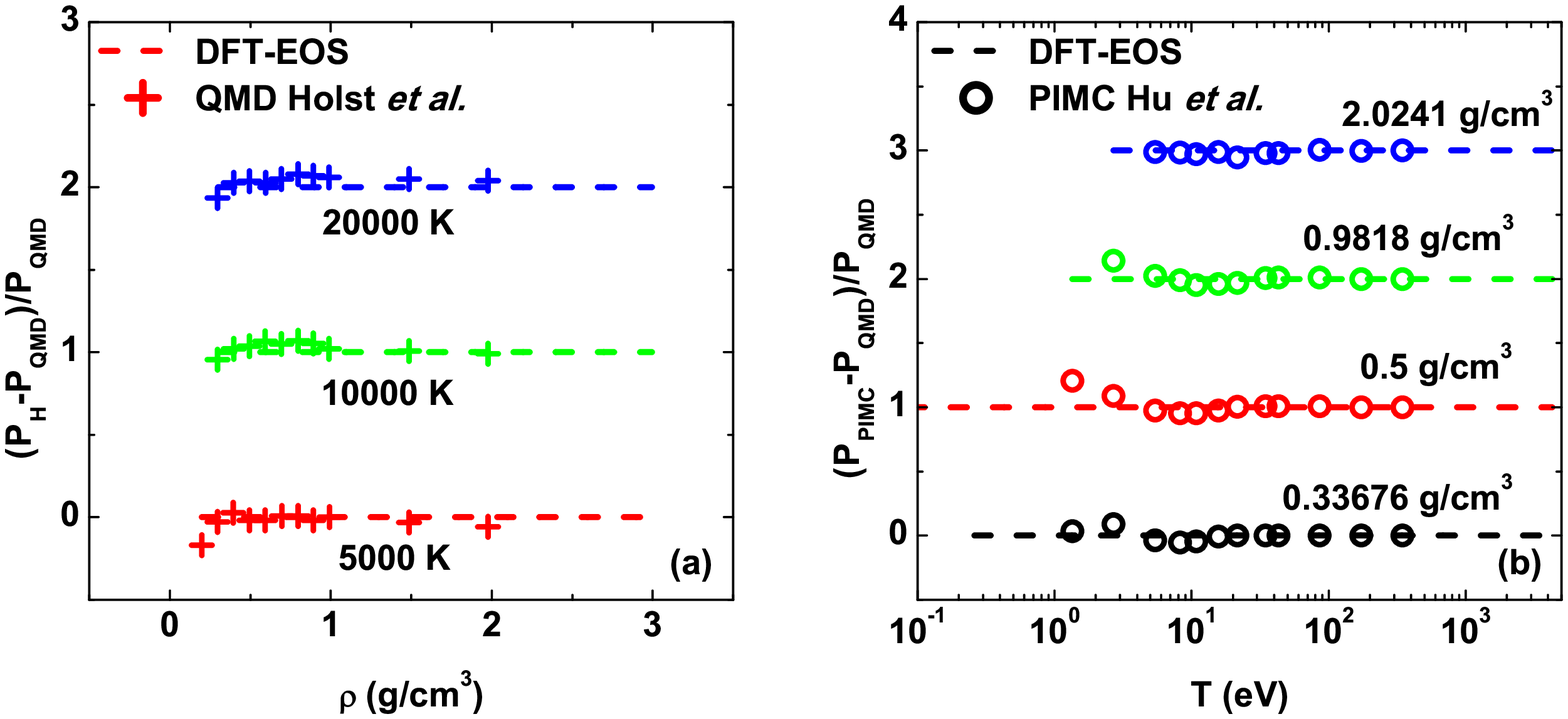}
\caption{(Color online) (a) Pressure difference as a function of
density, $P_{\text{H}}$ denotes data from Holst \emph{et al.}
\cite{Holst2008}; (b) Pressure difference as a function of
temperature, $P_{\text{PIMC}}$ are obtained from Hu \emph{et al.}
\cite{Hu2011}. Each curve has been shifted by 1.0 from the
previous one for clarity.}\label{fig_mid_P}
\end{figure}

In this section, the present DFT-EOS data have been systemically compared
with previous theoretical predictions. At densities from
$\sim10^{-3}$ g/cm$^{3}$ to $\sim10^{-1}$ g/cm$^{3}$, we have shown
the pressure and internal energy difference between our DFT-EOS and
those obtained by PIMC simulations (see Fig. \ref{fig_expanded}).
The results indicate a maximum of 7\% difference for the pressure
and 15\% for the energy over the temperatures we explore. At
temperatures above $\sim$ 30 eV, the distinction can be viewed as
small as negligible between the two methods. In the warm dense
regime, which is highlighted at densities between 0.2 and 3.0
g/cm$^{3}$, very good agreement has been found between our DFT-EOS
data and those fitted QMD results at the temperature domain 2000 $\sim$
10000 K (left panel in Fig. \ref{fig_E1}). At temperature beyond 10000 K, results from PIMC simulations by Militzer \emph{et
al.} \cite{Militzer2000}, Chabrier Model \cite{Chabrier1998}, and
QMD simulations \cite{Holst2008,Caillabet2011} have been plotted in the
right panel in Fig. \ref{fig_E1}. The PIMC method is suitable for investigating many-body quantum
systems at high temperatures. In this method, electrons and ions are
treated on equal footing as paths. The model of Chabrier and
Potekhin considers a fully ionized plasma, which is reliable at high
temperature and low density region. As shown in Fig.
\ref{fig_E1}, the present results are in accordance with those
numerical simulations and theoretical models.

\begin{figure}[!ht]
\includegraphics[height=6.0cm]{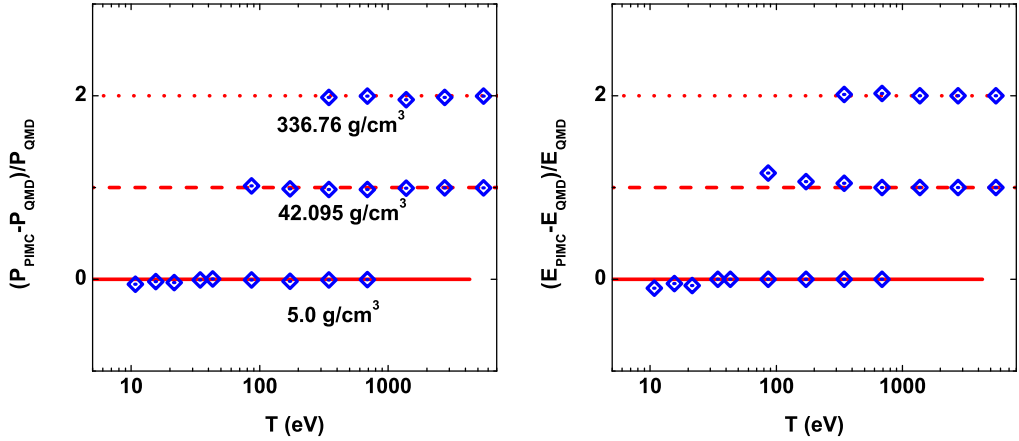}
\caption{(Color online) DFT-EOS is compared with PIMC data
\cite{Hu2011} in the dense plasma region. The present DFT-EOS
results are shown as red lines, and blue open diamonds denote PIMC
data. Each curve has been shifted by 1.0 from the previous one for
clarity.}\label{fig_EOS_dense}
\end{figure}

The isotherms of the pressure have been observed to show a
systematic behavior in terms of the density and temperature (see
Fig. \ref{fig_mid_P}). In this region, we do not find any signs
for $(\frac{\partial P}{\partial V})_{T}>0$, which would indicate
another first-order phase transition (the so-called PPT). PPT is usually considered in chemical models such
as fluid variational theory \cite{Juranek2002} or
liquid state perturbation theory \cite{Saumon1995}. In these
chemical models, minimization of the free energy for a mixture
consists of atoms, molecules, and plasma in equilibrium. Relations
between different particles are described by effective potentials.
As we explore to a higher temperature region (right panels in Fig.
\ref{fig_mid_P} and Fig. \ref{fig_EOS_dense}), PIMC data by Hu
\emph{et al.} \cite{Hu2011} are shown for comparison. It is
clearly indicated that the pressure given by DFT-EOS is in good
agreement with PIMC calculations up to $\sim10^{7}$ K for the
densities concerned, and this agreement extends toward lower
temperatures when the density decreases.

\section{Conclusion}

In summary, we have constructed a wide-range DFT-EOS by means of FPMD
and OFMD simulations. After building short cutoff radius Columbic
potential, we have the ability to explore the EOS into ultra-dense
region. The present DFT-EOS is valid at densities from
$9.82\times10^{-4}$ to $1.347\times10^{3}$ g/cm$^{3}$ with the
temperature up to $5\times10^{7}$ K. Available experimental data
and theoretical models have been introduced to compare with current
DFT-EOS. We have found good agreement between our results and those
data probed by gas gun, chemical explosive, and magnetic driven
plate flyer experiments, which indicate a maximum compression
ratio of 4.5 around 40 GPa. At higher pressures, our data show
stiff behavior and validates the high power laser experiments with
quartz standard. The principal Hugoniot curve is also accordant
with previous QMD simulation results. Agreement has also been found
between our DFT-EOS Hugoniot data and those obtained by
pre-compressed laser-driven shock wave experiments, which provide
visible ways to generate EOS in a broader density and temperature
regime. As density and
temperature enter into a denser and hotter regime, where
experimental detections are prohibited, the present results are
compared with those predicted by chemical model and PIMC
simulations. The present DFT-EOS covers typical states as can be
reached in ICF and will be applied in hydrodynamic simulations in
the future work.

\section{acknowledgement}
This work was supported by NSFC under Grants No. 11275032, No.
11005012 and No. 51071032, by the National Basic Security Research
Program of China, and by the National High-Tech ICF Committee of
China.


\begin{thebibliography}{99}
\bibitem {timestep}The time steps have been taken
as $\Delta t=a/20\sqrt{k_{B}T/m_{H}}$, where $a=(3/4\pi n_{i})^{1/3}$ is
the ionic sphere radius ($n_{i}$ is the ionic number density), $k_{B}T$
presents the kinetic energy, and $m_{H}$ is the ionic hydrogen.
\end{thebibliography}
\end{document}